# The Relativity of Simultaneity: An Analysis Based on the Properties of Electromagnetic Waves

R. Wayne

*Department of Plant Biology, Cornell University, Ithaca, NY 14853 USA*

The determination of whether two distant events are simultaneous depends on the velocity of the observer. This velocity dependence is typically explained in terms of the relativity of space and time in a counterintuitive manner by the Special Theory of Relativity. In this paper, I describe a straightforward and intuitive way to explain the velocity dependence of simultaneity in terms of velocity-dependent changes in the spatial (**k**, $\lambda$) and temporal ($\omega$, $v$) characteristics of electromagnetic waves that result from the Doppler effect. Since, for any solution to a wave equation, the angular wave vector (**k**) and distance vector (**r**) as well as the angular frequency ($\omega$) and time ($t$) are complementary pairs (**k** · **r**) and ($\omega t$), it is only a matter of taste which members of the pairs (**k**, $\omega$) or (**r**, $t$) one assumes to depend on the relative velocity of the source and observer. Einstein chose **r** and $t$ and I chose **k** and $\omega$. I present this electromagnetic wave approach to understanding the velocity dependence of simultaneity as a physically realistic alternative to Einstein's Special Theory of Relativity.

**Introduction**

In the late 1800s, the introduction of fast moving trains and high-speed telegraphic communication forced a rethinking of the nature of space and time. In terms of society, this rethinking resulted in the elimination of local time and the adoption of standard time and time zones. The introduction of standard time allowed passengers traveling long distances to make connections easily between trains originating at distant stations [1], and telegraphers to be at the station at a specific time to send a message to or receive a message from a distant place [2]. Perhaps these technological changes caused Einstein to think twice about the nature of time [3]. Einstein could simplify a variety of scientific problems in the fields of dynamics, electromagnetism and optics by postulating that time itself was relative and depended on the velocity of the observer relative to the system observed [3]. Einstein began his rethinking of the nature of time by considering the concept of simultaneity and the methods used to synchronize clocks. He realized that the reckoning of simultaneity depended on the velocity of the observer. Einstein's rethinking resulted in the Special Theory of Relativity that states that the velocity dependence of simultaneity is a consequence of the relativity of space and time. I suggest that the velocity dependence of simultaneity and time can be explained better by the velocity-dependent characteristics of electromagnetic waves, as exemplified by the Doppler effect. Perhaps the ubiquity in the twenty-first century of Doppler radar used in weather forecasting [4], Doppler ultrasound used in medical diagnosis [5] and the roadside Doppler radar used by police influenced me to choose the temporal and spatial characteristics of electromagnetic waves instead of time and space as the physically relevant, velocity-dependent variables that are capable of accounting for the relativity of simultaneity.

It is commonplace that the determination of whether two distant events are simultaneous or not is relative and depends on the position of the observer (Figure 1). For example, when an observer is standing midway between two identical lamps, both of which are in the same inertial frame as the observer, the observer would say that the two lamps came on simultaneously if the light from the two lamps reached him or her at the same time. For this observer, the duration of time between when the first and second lamp came on would be zero. However, due to the finite speed of light [6, 7], a second observer, who is closer to the lamp on the left would not see the two lamps come on simultaneously–but would see the lamp on the left come on before the lamp on the right and would measure a finite duration of time between when the first and second lamp came on. A third observer in the same inertial frame, who is closer to the lamp on the right would see the lamp on the right come on before the lamp on the left and would also measure a finite duration of time between when the first and second lamp came on. The sequence of events clocked by the third observer would be the reverse of the sequence of events clocked by the second observer. These examples show that without making any assumptions other than that each observer has an identical clock; the measurement of the time interval between two events is relative. Of course, if each observer knew his or her position relative to the two lamps and the speed of light (c) through the air, then using the following relationship:

$$time\ interval = \frac{length\ of\ light\ path}{speed\ of\ light}, \quad (1)$$

all three observers would be able to agree when the two lamps turned on. The resolution of this problem requires nothing more than a physically-meaningful theory of measurement that takes into consideration the finite speed of light.

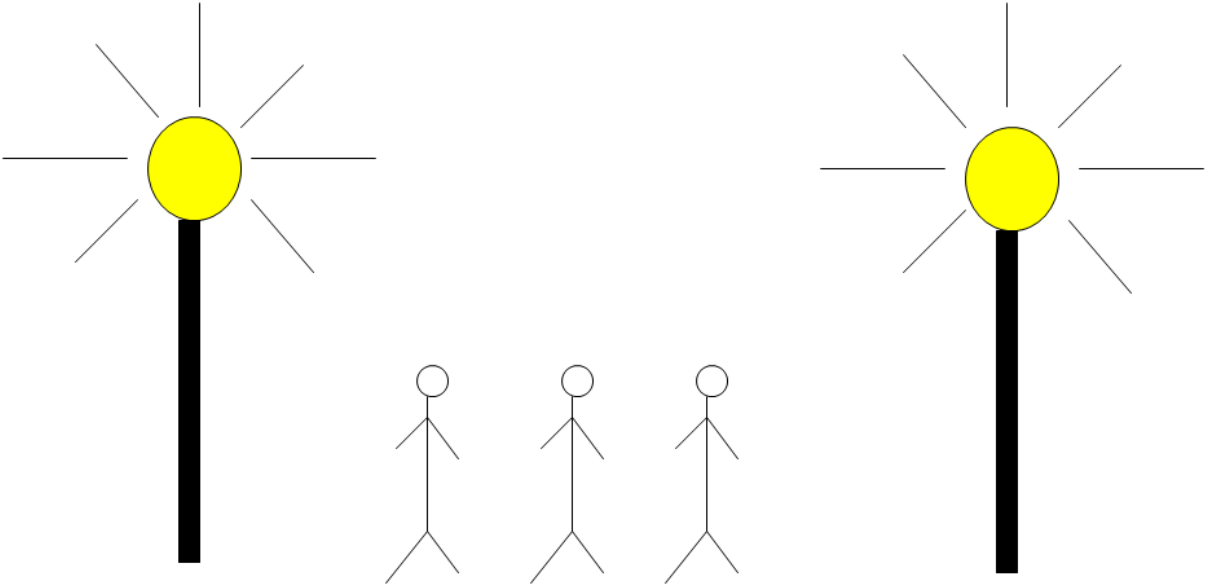

Fig. 1. The relativity of simultaneity for observers and events (lamps turning on) in the same inertial frame. When the observer in the middle sees the two lamps come on simultaneously, the observer on the left sees the lamp on the left come on first, while the observer on the right sees the lamp on the right come on first.

With great insight, Einstein realized that the reckoning of whether two events were simultaneous or not depended on the observer's velocity ($v$) relative to the two identical lamps [8, 9, 10], in addition to his or her distance ($L$) from them. Imagine two observers, as Comstock [11] and Einstein [9] did, standing midway between a lamp mounted on the front of a railroad car and a lamp mounted on the back (Figure 2). Imagine that one observer is on the railroad car and the other observer is on the platform. The observer on the moving railroad car, at rest with respect to the lamps, would see the lamps come on simultaneously as predicted by equation 1. However, the observer on the platform, moving at relative velocity $v$ toward the lamp at the back of the railroad car and at relative velocity $v$ away from the lamp at the front of the railroad car, would see the lamp at the back of the railroad car come on before the lamp at the front of the railroad car came on. Even though both observers were midway between the two lamps, the observer on the railroad car would have seen the lamps come on simultaneously, while the observer on the platform would not have.

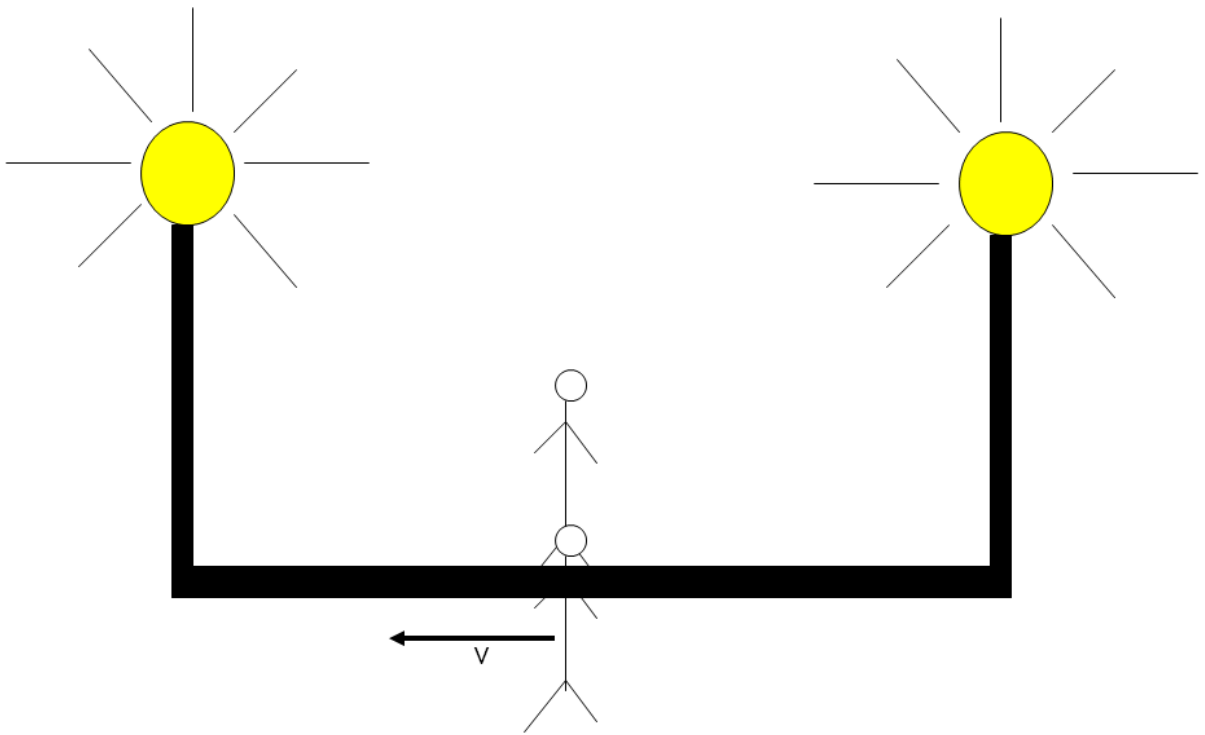

Fig. 2. The relativity of simultaneity for observers, standing midway between the two lamps, in two different inertial frames. One observer is in a railroad car at rest with respect to a lamp mounted on the back of the railroad car and another lamp mounted on the front of the railroad car. The other observer is standing on the platform moving backwards at velocity $v$ relative to the train. The observer on the railroad car sees the two lamps come on simultaneously, while the observer on the platform sees the lamp on the back of the railroad car come on before the lamp on the front of the railroad car comes on.

According to the Special Theory of Relativity, the inability of two observers, in different inertial frames, to agree on when two events occurred, and, whether they were simultaneous events, is a consequence of the relativity of time. That is, the Special Theory of Relativity contends that time itself is relative, and consequently, its measurement depends on the velocity of the observer. Quantitatively, the observer on the platform who is moving backwards relative to the train would see the light come on from the lamp at the back of the railroad car $dt_{observer-back}$ seconds after it was emitted and would see the light come on from the lamp at the front of the railroad car $dt_{observer-front}$ seconds after it was emitted. This is described in the following equations:

$$dt_{observer-back} = \frac{\frac{L}{c} - \frac{vL}{c^2}}{\sqrt{1 - \frac{v^2}{c^2}}} \tag{2}$$

$$dt_{observer-front} = \frac{\frac{L}{c} + \frac{vL}{c^2}}{\sqrt{1 - \frac{v^2}{c^2}}} \tag{3}$$

The duration of time (Δ) between when the two lamps come on depends on the velocity of the observer relative to the lamps and is given by the Lorentz transformation for time:

$$\Delta = dt_{observer-front} - dt_{observer-back} = \frac{2\frac{vL}{c^2}}{\sqrt{1 - \frac{v^2}{c^2}}} \tag{4}$$

where the duration of time between when the two lamps come on depends on the relativity of time itself. The relativity of time is given quantitatively by the time dilation factor, $\gamma = \frac{1}{\sqrt{1 - \frac{v^2}{c^2}}}$. The duration of time between when the lamps at the front and back of the railroad car come on vanishes for an observer midway between the two lamps when $v = 0$. The "two-way" duration, which is a ubiquitous quantity in the Special Theory of Relativity, can be obtained by taking the average of the "one-way" durations:

$$dt_{two-way} = \frac{1}{2}(dt_{observer-front} + dt_{observer-back}) = \frac{\frac{L}{c}}{\sqrt{1 - \frac{v^2}{c^2}}} \tag{5}$$

The relativity of simultaneity is often illustrated with a Minkowski space-time diagram [12]. Figure 3 shows the reckoning of an observer (a) who is stationary with respect to the lamps at the front and back of the railroad car, and the reckoning of an observer (b) who is moving with velocity $v$ toward the lamp at the back of the railroad car. While the concept that the velocity-dependent relativity of simultaneity is a consequence of the fundamental nature and relativity of time is widely and deeply accepted by modern physicists, I would like to offer an alternative explanation that is based on the primacy of the Doppler effect, which takes into consideration the velocity-dependent temporal and spatial characteristics of electromagnetic waves, including their wavelength ($\lambda$), their frequency ($v$), their angular wave number ($k$), and their angular frequency ($\omega$), in addition to their speed.

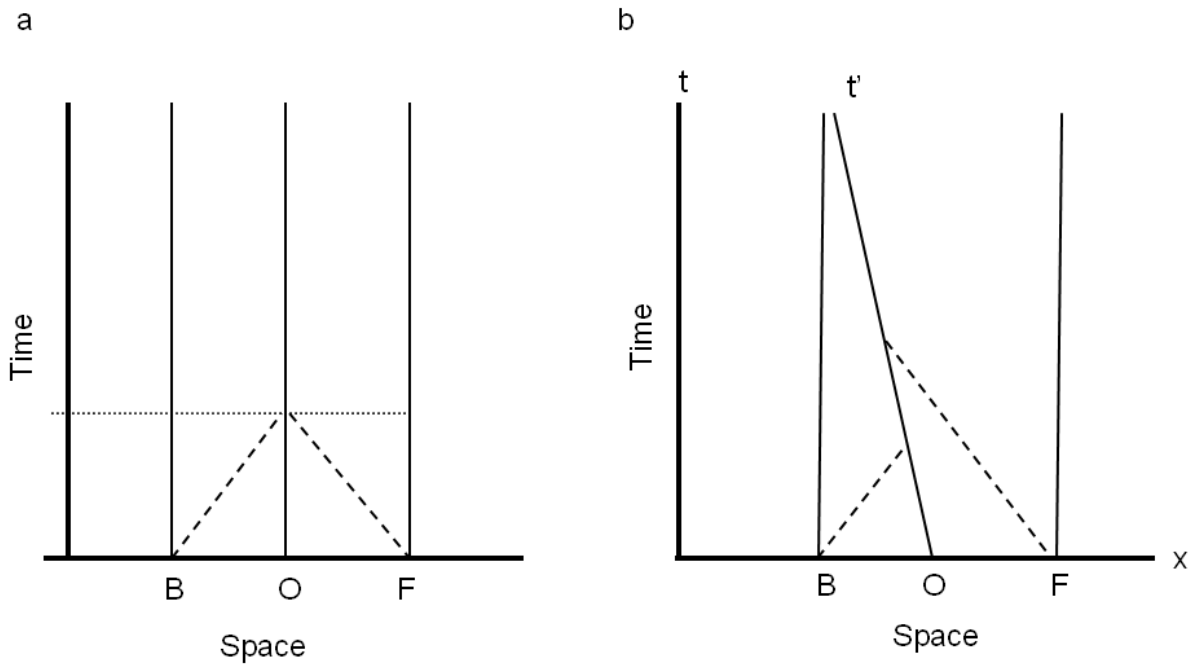

Fig. 3. The relativity of simultaneity. a. A Minkowski space time diagram of the observer (O) in the railroad car midway between the lamp mounted on the back (B) of the railroad car and the lamp mounted on the front (F) of the railroad car. This observer sees the two identical lights come on simultaneously. b. A Minkowski space time diagram of the observer (O) on the platform moving backwards at velocity $v$ relative to the railroad car. When this observer is midway between the lamp mounted on the back (B) of the railroad car and the lamp mounted on the front (F) of the railroad car, he or she sees the lamp on the back of the railroad car come on before the lamp on the front of the railroad car comes on. t represents the time in the frame of reference of the lamps and t' represents time in the frame of reference of the observer on the platform.

**Results and Discussion**

Einstein tried to reformulate Maxwell's equations in a way that would take into consideration two inertial frames moving relative to each other at velocity *v*, but his attempts were unsuccessful [13]. Consequently, he assumed that Maxwell's wave equation with its single explicit velocity (c) was one of the laws of physics that was valid in all inertial frames and, as a result, the speed of light was independent of the relative velocity of the source and the observer when they were in two different inertial frames. I have reformulated Maxwell's wave equation so that it takes into consideration the changes in the temporal and spatial characteristics of electromagnetic waves observed when there is relative motion between the inertial frame that includes the source and the inertial frame that includes the observer. The new relativistic wave equation presented here is form-invariant to the second order in all inertial frames. My reformulation of Maxwell's wave equation is based on the primacy of the Doppler effect, which is experienced by all waves, as opposed to the primacy of the relativity of space and time. Since, for any solution to the second order wave equation in the form of $\Psi = \Psi_o e^{i(\mathbf{k} \cdot \mathbf{r} - \omega t)}$, the angular wave vector ($\mathbf{k}$) and distance ($\mathbf{r}$) as well as the angular frequency ($\omega$) and time ($t$) are complementary pairs ($\mathbf{k} \cdot \mathbf{r}$) and ($\omega t$), it is only a matter of taste which members of the pairs ($\mathbf{k}$, $\omega$) or ($\mathbf{r}$, $t$) one assumes to depend on the relative velocity of the source and observer. Einstein chose $\mathbf{r}$ and $t$ and I chose $\mathbf{k}$ and $\omega$. The Doppler-based relativistic wave equation is given by the following equivalent forms:

$$\frac{\partial^2 \Psi}{\partial t^2} = cc' \frac{\sqrt{c + v\cos\theta}}{\sqrt{c - v\cos\theta}} \nabla^2 \Psi \tag{6}$$

$$\frac{\partial^2 \Psi}{\partial t^2} = cc' \frac{\sqrt{1 + \frac{v\cos\theta}{c}}}{\sqrt{1 - \frac{v\cos\theta}{c}}} \nabla^2 \Psi \tag{7}$$

$$\frac{\partial^2 \Psi}{\partial t^2} = cc' \frac{1 + \frac{v}{c}\cos\theta}{\sqrt{1 - \frac{v^2 \cos^2\theta}{c^2}}} \nabla^2 \Psi \tag{8}$$

where *v* is the magnitude of the relative velocity of the source and observer; $\theta$ is the angle subtending the velocity vector of the source or the observer and the wave vector originating at the source and pointing toward the observer assuming the rotation is counterclockwise; c is the speed of light through the vacuum and is equal to the square root of the reciprocal of the product of the electric permittivity ($\varepsilon_o$) and the magnetic permeability ($\mu_o$) of the vacuum; and *c'* is the ratio of the angular frequency ($\omega_{source}$) of the source in its inertial frame to the angular wave number ($k_{observer}$) observed in any inertial frame. When the velocity vector and the angular wave vector are parallel and antiparallel, $\theta = 0$, $\cos\theta = 1$ and $\theta = \pi$ radians, $\cos\theta = -1$, respectively. Solving the relativistic wave equation given above for the speed of the wave (c = r/t) results in the following relativistic dispersion relation (see Appendix A):

$$c = \frac{\omega_{source}}{k_{observer}} \frac{\sqrt{1 + \frac{v \cos \theta}{c}}}{\sqrt{1 - \frac{v \cos \theta}{c}}} = 2.99 \times 10^8 \text{ m/s} \tag{9}$$

When $v = 0$, the source and the observer are in the same inertial frame and $\omega_{source} = k_{source}c$. After replacing $\omega_{source}$ with $k_{source}c$, the above equation transforms into a perspicuous relativistic equation that describes the new relativistic Doppler effect:

$$k_{observer} = k_{source} \frac{\sqrt{1 + \frac{v \cos \theta}{c}}}{\sqrt{1 - \frac{v \cos \theta}{c}}} \tag{10}$$

$$k_{observer} = k_{source} \frac{1 + \frac{v}{c} \cos \theta}{\sqrt{1 - \frac{v^2 \cos^2 \theta}{c^2}}} \tag{11}$$

The above equation that describes the new relativistic Doppler effect differs from Einstein's relativistic Doppler effect equation by having a cosine term in both the numerator and the denominator. The cosine term describes the dependence of the first-order and second-order velocity-dependent spatial and temporal properties of electromagnetic waves on the component of the velocity relative to the angular wave vector. The two cosine terms ensure that the effective velocity between the source and the observer is completely relative and depends only on the source and the observer. By contrast, Einstein's equation for the relativistic Doppler effect is:

$$k_{observer} = k_{source} \frac{1 + \frac{v}{c} \cos \theta}{\sqrt{1 - \frac{v^2}{c^2}}} \tag{12}$$

In Einstein's formulation, the first-order velocity-dependent spatial and temporal properties of electromagnetic waves depends on the component of the velocity parallel to the angular wave vector. By contrast, the second-order velocity-dependent spatial and temporal properties of waves depends on the speed as opposed to the velocity. In order to leave the cosine term out of the denominator, Einstein [8] had to assume that the velocity applies to a situation where there is an "infinitely distant source of light" and consequently $\cos^2 \theta$ is equal to unity. This assumption limits the applicability of Einstein's relativistic Doppler effect equation. The velocity in the denominator is not relative in the true sense of the word since it cannot be completely determined solely by an observer localized at a given coordinate when $\cos^2 \theta$ is not equal to unity but only by an omniscient observer.

Qualitatively, the Doppler effect [14] characterizes the changes that occur in the temporal and spatial characteristics of a wave as a function of the relative velocity of the source and the observer. Quantitatively, the magnitude of the predicted Doppler effect depends on the relativistic transformation used to describe the relationship between two inertial frames. Doppler, himself, utilized the Galilean transformation $(1 + \frac{v \cos \theta}{c})$, the only transformation available at the time, to describe the velocity-dependent changes in the temporal and spatial characteristics of light and sound waves that occur when the source and observer are in two different inertial frames. Einstein [8] modified the Galilean transformation with the newly accessible and dimensionless Lorentz factor ($\frac{1+\frac{v\cos\theta}{c}}{\sqrt{1-\frac{v^2}{c^2}}}$), in order to describe the velocity-dependent changes in the spatial and temporal characteristics of light waves that occur when the source and observer are in two different inertial frames. Einstein's formula, but not that proposed by Doppler, was validated for light waves when $\theta$ was equal to 0 and $\pi$ by the experiments done by Ives and Stillwell [15, 16]. The formula I have proposed for the Doppler effect, which is also consistent with the Ives-Stillwell experiments, makes use of both the Galilean transformation and a Lorentz-like factor ($\frac{1+\frac{v\cos\theta}{c}}{\sqrt{1-\frac{v^2\cos^2\theta}{c^2}}}$). The physical justification of my transformation is its ability to model the results of the Ives-Stillwell experiments. A mathematical justification is given in Appendix B. In Doppler's, Einstein's and my formulations, when the source is stationary, an approaching observer ($\theta$ = 0) encounters more waves per unit time, while a receding observer ($\theta$ = $\pi$) encounters fewer waves per unit time; and, when the observer is stationary, a receding source ($\theta$ = $\pi$) produces fewer waves per unit time at the position of the observer, while an approaching source ($\theta$ = 0) produces more waves per unit time at the position of the observer. The net result of the Doppler effect is an increase in $k$, $\omega$ and $v$ and a decrease in $\lambda$ reckoned by the observer when the source and observer move closer together and a decrease in $k$, $\omega$ and $v$ and an increase in $\lambda$ reckoned by the observer when the source and the observer move apart.

    The experimental observations of Ives and Stillwell [15] on the effect of velocity on the displacement of the spectral lines of hydrogen ions confirm the utility and validity of using the new relativistic wave equation. However, the predictions of the new relativistic wave equation differ in other ways from the predictions of the Special Theory of Relativity. For example, the Special Theory of Relativity [8, 17] predicts the existence of a transverse Doppler shift exactly perpendicular to the velocity of an inertial particle, while the new relativistic wave equation does not. Since it is difficult to measure the transverse Doppler effect in an inertial system [18], experiments approximate the transverse Doppler shift by averaging the forward and backward longitudinal Doppler shifts [15, 19]. Both the Special Theory of Relativity and the new relativistic wave equation presented above predict that averaging the forward and backward longitudinal Doppler-shifted light will give the Lorentz factor also known as the "time dilation" factor as observed in such experiments. The fact that Ives [20, 21, 22, 23] never interpreted his own results as a confirmation of the Special Theory of Relativity provides a reason to think twice about alternative explanations. Spectroscopic techniques that take into consideration the angular dependence of the anisotropy [24, 25, 26, 27, 28, 29] could be used to test the quantitatively-different predictions of Einstein's relativistic Doppler effect equation and the new relativistic Doppler effect equation presented here.

    If the lamps on the front and back of a train are identical and emit light with an angular wave number of $k_{source}$, as a result of the Doppler effect, the angular wave number of the light emitted by the lamp at the back of the railroad car would appear to the observer on the platform to have a greater angular wave number ($k = \frac{2\pi}{\lambda}$) than the light emitted by the lamp at the front of

the railroad car. The velocity dependence of the angular wave number of the light seen coming from the lamps on the back and front of the railroad car reckoned by an observer on the platform is given by the following equation:

$$k_{observer} = k_{source} \frac{1 + \frac{v}{c}\cos\theta}{\sqrt{1 - \frac{v^2 \cos^2\theta}{c^2}}} \tag{13}$$

In the case shown in figure 4, where $\theta$ is equal to $\frac{7\pi}{4}$ for light coming from the lamp on the back of the train and $\frac{5\pi}{4}$ for light coming from the lamp on the front of the train, we get:

$$k_{observer\text{-}back} = k_{source} \frac{1 + 0.707\frac{v}{c}}{\sqrt{1 - \frac{v^2}{2c^2}}} \tag{14}$$

$$k_{observer\text{-}front} = k_{source} \frac{1 - 0.707\frac{v}{c}}{\sqrt{1 - \frac{v^2}{2c^2}}} \tag{15}$$

Since the momentum of photons is given by $\hbar k$, the observer on the platform would also reckon the momentum of the photons emitted by the lamp on the back of the railroad car as being greater than the momentum of the photons being emitted by the lamp on the front of the railroad car. Similarly, if the lamps on the front and back of a train are identical and emit light with an angular frequency of $\omega_{source}$, as a result of the Doppler effect, the angular frequency of the light emitted by the lamp at the back of the railroad car would appear to the observer on the platform to have a greater angular frequency than the light emitted by the lamp at the front of the railroad car. The velocity dependence of the angular frequency of the light seen coming from the lamps on the back and front of the railroad car reckoned by an observer on the platform is given by the following equation:

$$\omega_{observer} = \omega_{source} \frac{1 + \frac{v}{c}\cos\theta}{\sqrt{1 - \frac{v^2 \cos^2\theta}{c^2}}} \tag{16}$$

In the case shown in figure 4, where $\theta$ is equal to $\frac{7\pi}{4}$ for light coming from the lamp on the back of the train and $\frac{5\pi}{4}$ for light coming from the lamp on the front of the train, we get:

$$\omega_{observer\text{-}back} = \omega_{source} \frac{1 + 0.707\frac{v}{c}}{\sqrt{1 - \frac{v^2}{2c^2}}} \tag{17}$$

$$\omega_{observer\text{-}front} = \omega_{source} \frac{1 - 0.707\frac{v}{c}}{\sqrt{1 - \frac{v^2}{2c^2}}} \tag{18}$$

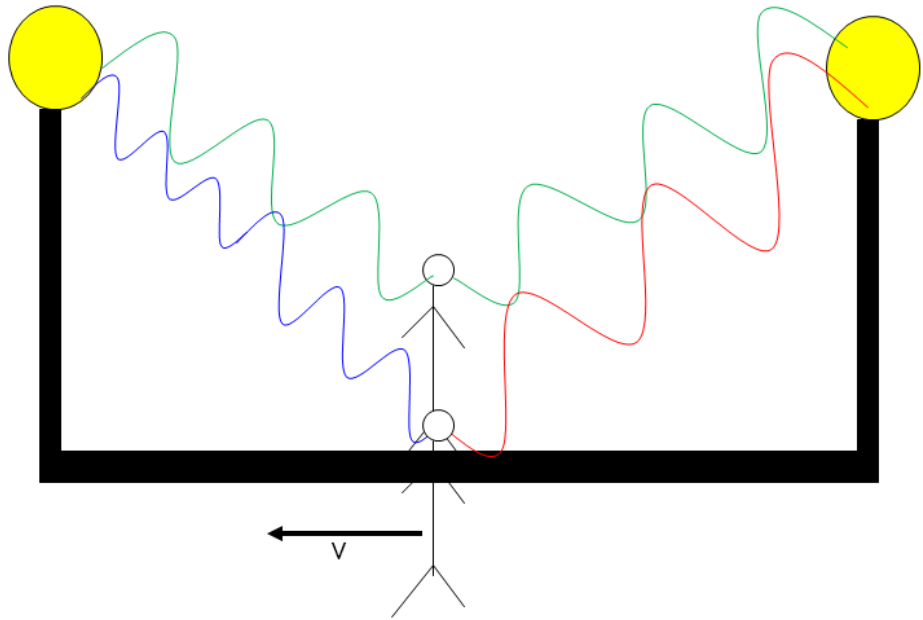

Fig. 4. The observer in the railroad car midway, between the lamps on the back and front of the railroad car, sees the two identical lights come on simultaneously. As a consequence of the Doppler effect, the observer on the platform moving backwards at velocity *v* relative to the railroad car, and who is midway between the lamp mounted on the back of the railroad car and the lamp mounted on the front of the railroad car, sees the light emitted by the lamp on the back ($\theta = \frac{7\pi}{4}$) of the railroad car as being blue-shifted and the light emitted from the lamp at the front ($\theta = \frac{5\pi}{4}$) of the car as being red-shifted. While the velocities of the blue-shifted and red-shifted lights are the same and equal to c, the amplitude and energy of the blue-shifted wave arrives at the observer before the amplitude and energy of the red-shifted wave. Consequently, the observer on the platform does not see the two lamps come on simultaneously.

     Since the energy of a photon is given by $\hbar\omega$, the observer on the platform would also reckon the energy of the photons emitted by the lamp on the back of the railroad car as being greater than the energy of the photons being emitted by the lamp on the front of the railroad car.

     Even if the lamps on the front and back of a train are identical and emit light with a wavelength of $\lambda_{source}$, as a result of the Doppler effect, the wavelength of the light emitted by the

lamp at the back of the railroad car would appear to the observer on the platform to be shorter than the light emitted by the lamp at the front of the railroad car.

The velocity dependence of the wavelength of the light seen coming from the lamps on the back and front of the railroad car reckoned by an observer on the platform is given by the following equation:

$$\lambda_{observer} = \lambda_{source} \frac{\sqrt{1 - \frac{v^2 \cos^2 \theta}{c^2}}}{1 + \frac{v}{c}\cos\theta} = \lambda_{source} \frac{1 - \frac{v}{c}\cos\theta}{\sqrt{1 - \frac{v^2 \cos^2 \theta}{c^2}}} \quad (19)$$

In the case shown in figure 4, where $\theta$ is equal to $\frac{7\pi}{4}$ for light coming from the lamp on the back of the train and $\frac{5\pi}{4}$ for light coming from the lamp on the front of the train, we get:

$$\lambda_{observer\text{-}back} = \lambda_{source} \frac{1 - 0.707\frac{v}{c}}{\sqrt{1 - \frac{v^2}{2c^2}}} \quad (20)$$

$$\lambda_{observer\text{-}front} = \lambda_{source} \frac{1 + 0.707\frac{v}{c}}{\sqrt{1 - \frac{v^2}{2c^2}}} \quad (21)$$

Since the initial peak amplitude of a wave would reach an observer $\frac{\lambda}{4}$ after the leading edge did, and since, to an observer on the platform, midway between the two lamps, the wavelength of the light originating from the lamp at the back of the railroad car would be shorter than the wavelength of the light originating from the lamp at the front of the railroad car, the observer on the platform would observe the lamp on the back of the railroad car come on before the lamp on the front of the railroad car. *Note that while the phases of the leading edges of the electromagnetic waves reaching the observer on the platform would be the same, the phases of the peak amplitudes would not.*

Furthermore, if the lamps on the front and back of a train are identical and emit light with a frequency of $v_{source}$, as a result of the Doppler effect, the frequency of the light emitted by the lamp at the back of the railroad car would appear to the observer on the platform to have a greater frequency than the light emitted by the lamp at the front of the railroad car. The velocity dependence of the frequency of the light seen coming from the lamps on the back and front of the railroad car reckoned by an observer on the platform is given by the following equation:

$$v_{observer} = v_{source} \frac{1 + \frac{v}{c}\cos\theta}{\sqrt{1 - \frac{v^2 \cos^2 \theta}{c^2}}} \quad (22)$$

In the case shown in figure 4, where $\theta$ is equal to $\frac{7\pi}{4}$ for light coming from the lamp on the back of the train and $\frac{5\pi}{4}$ for light coming from the lamp on the front of the train, we get:

$$v_{observer\text{-}back} = v_{source} \frac{1 + 0.707\frac{v}{c}}{\sqrt{1 - \frac{v^2}{2c^2}}} \qquad (23)$$

$$v_{observer\text{-}front} = v_{source} \frac{1 - 0.707\frac{v}{c}}{\sqrt{1 - \frac{v^2}{2c^2}}} \qquad (24)$$

Since the frequency of a wave is a measure of the rate of energy, momentum and information transfer, and since, to an observer on the platform, midway between the two lamps, the frequency of the light originating from the lamp at the back of the railroad car would be higher than the frequency of the light originating from the lamp at the front of the railroad car, the observer on the platform would detect the energy, momentum and information coming from the lamp on the back of the railroad car before he or she detected the energy, momentum and information coming from the lamp on the front of the railroad car. The relations described in equations 10-24 hold even when each lamp is reduced to a single vibrating atom acting as a clock.

To an observer in the railroad car, at rest with respect to the lamps ($v = 0$), the light originating from the lamps on the back and the front of the train would be isotropic in terms of its angular wave number, angular frequency, wavelength and frequency, while the light reaching the observer on the platform would be anisotropic in terms of these wave characteristics (Figure 4). The quantitative difference in the angular dependence of the anisotropy predicted by the new relativistic Doppler effect equation presented here and Einstein's relativistic Doppler effect equation could be tested with spectroscopic techniques [24, 25, 26, 27, 28, 29].

To an observer on the railroad car who is at rest ($v = 0$) with respect to the lamps, the durations of time it would take the light emitted by lamps at the back and front of the railroad car to reach the observer would be symmetrical, while to the observer on the platform, the durations of time it would take the light, emitted by lamps at the back and front of the railroad car, to reach the observer would be asymmetrical. As a result of the Doppler effect, the duration of time it would take the light from the lamp at the back or the front of the railroad car to reach the observer on the platform moving at velocity $v$ relative to the train would be:

$$dt_{observer} = \frac{N}{v_{source}} \frac{\sqrt{1 - \frac{v^2 \cos^2\theta}{c^2}}}{1 + \frac{v}{c}\cos\theta} = \frac{L}{c} \frac{1 - \frac{v}{c}\cos\theta}{\sqrt{1 - \frac{v^2 \cos^2\theta}{c^2}}} \qquad (25)$$

where $N$ is the number of waves between the source and the observer and is equal to $\frac{L}{\lambda_{source}}$ and $\frac{N}{v_{source}} = \frac{L}{c}$. In the case shown in figure 4, where $\theta$ is equal to $\frac{7\pi}{4}$ for light coming from the lamp on the back of the train, we get:

$$dt_{observer\text{-back}} = \frac{L}{c} \frac{1 - 0.707\frac{v}{c}}{\sqrt{1 - \frac{v^2}{2c^2}}} \quad (26)$$

The duration of time it would take the light from the lamp at the front ($\theta = \frac{5\pi}{4}$) of the railroad car to reach the observer on the platform would be:

$$dt_{observer\text{-front}} = \frac{L}{c} \frac{1 + 0.707\frac{v}{c}}{\sqrt{1 - \frac{v^2}{2c^2}}} \quad (27)$$

The difference ($\Delta$) in the times it would take for light from the lamps at the front and back of the railroad car to reach the observer on the platform would be:

$$\Delta = dt_{observer\text{-front}} - dt_{observer\text{-back}} = \frac{L}{c} \frac{\frac{2v\cos\theta L}{c^2}}{\sqrt{1 - \frac{v^2 \cos^2\theta}{c^2}}} \quad (28)$$

The above equation reduces to equation 4 when $\theta = 0$. When $v$ equals zero, the Doppler effect vanishes and an observer midway between two events would reckon those events to occur simultaneously. However, as the relative velocity ($v$) of the inertial frame of the source and the inertial frame of the observer approaches c, the difference ($\Delta$) in time between the two events gets larger and larger and approaches infinity. The "two-way" duration of the Special Theory of Relativity, which is given in equation 5, and is a necessary device for synchronizing clocks, can also be obtained by letting $\theta = 0$ or $\pi$ and taking the average of the two "one-way" durations derived from the new relativistic Doppler effect equation:

$$dt_{two\text{-}way} = \frac{1}{2}(dt_{observer-back} + dt_{observer-back}) = \frac{\frac{L}{c}}{\sqrt{1 - \frac{v^2}{c^2}}} \quad (29)$$

The new relativistic Doppler effect equation, which is a more general expression of the relativistic Doppler effect because it does not assume an infinitely-distant source, can account for the velocity-dependence of the reckoning of simultaneity as a limiting case. Moreover, the "two-way" duration of the Special Theory of Relativity results in a loss of the spatial and temporal information that is retained by using the new relativistic Doppler equation.

Another way of looking at the velocity-dependent asymmetry is to look at the slew rate ($\partial\Psi/\partial t$) of the electromagnetic waves emitted by the two lamps. The leading edges of the electromagnetic waves, which contain no momentum, energy or information, arrive from the back and the front of the railroad car simultaneously at the two inertial observers.

While, to the observer on the railroad car, at rest with respect to the lamps, the slew rate of the electromagnetic waves emitted by the lamps on the back and front of the railroad car are the same, to the observer on the platform, by contrast, the slew rate of the electromagnetic waves from the back of the railroad car is greater than the slew rate of the electromagnetic waves from the front of the railroad car. Thus the observer on the platform detects the amplitude, momentum, energy and information of the electromagnetic waves from the back of the railroad car before he or she detects these qualities of the electromagnetic waves from the front of the railroad car. Figure 5 shows the temporal dependence of the wave-mediated transport of information in the form of amplitude and energy to the observer on the platform from the front and back of the railroad car. It also illustrates the Doppler effect-induced time lags between two waves with the same phase but different frequencies reckoned by an observer on the platform, midway between the lamp on the back of the railroad car and the lamp on the front of the railroad car, and moving with relative velocity $v$ towards the back of the train.

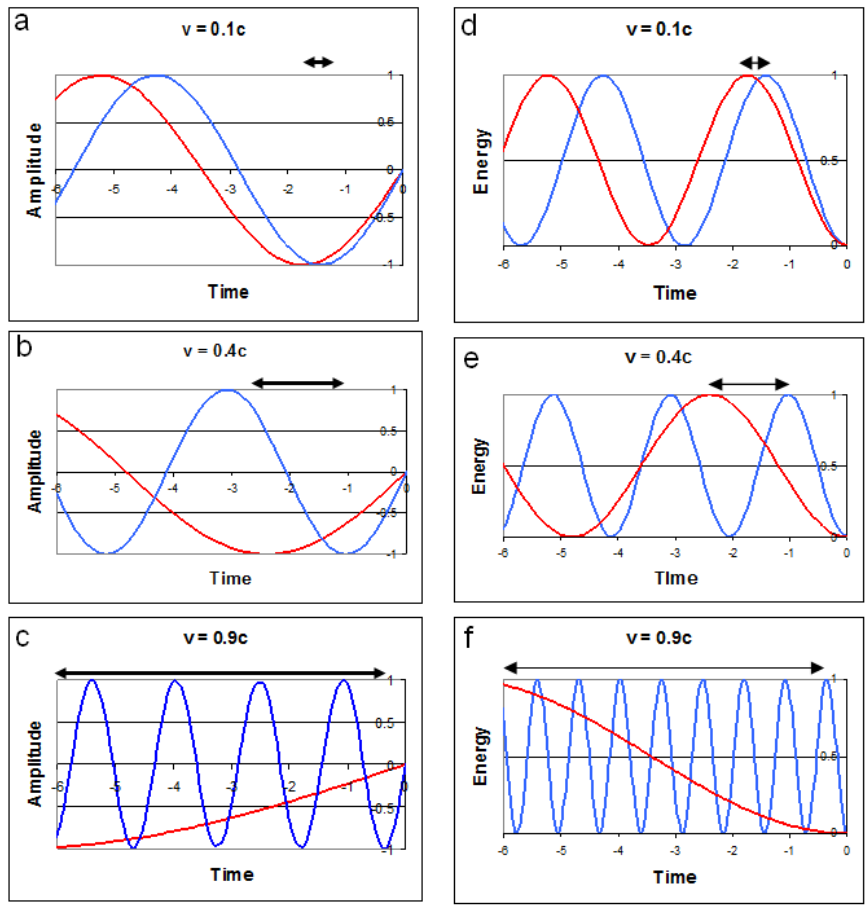

Fig. 5. The light wave (in blue) coming from the lamp at the back of the railroad car arrives at the observer on the platform at $t = 0$. The light wave coming from the lamp at the front (in red) of the railroad car arrives at the observer on the platform at $t = 0$. While the phases ($\alpha$) of the two waves (at $t = 0$) are the same, there is a time lag, introduced by the Doppler effect, between the peak amplitude of the wave coming from the lamp on the back of the railroad car and the peak amplitude of the wave coming from the lamp on the front of the railroad car. The duration of the Doppler effect-induced time lag is represented by the double arrow (↔). The time lag between the wave emitted by the lamp on the front of the railroad car and the wave emitted by the lamp on the back of the railroad car is presented in terms of amplitude (a,b,c) and energy (d,e,f) for observers moving relative to the train at velocities of 0.1c, 0.4c and 0.9c. $\theta$ is assumed to be equal to $\pi$ for the light coming from the back of the train and equal to 0 for the light coming from the front of the train.

If we consider the square of the amplitude of the electromagnetic waves emitted from the lamps to be proportional to the probability of detecting information-bearing photons, then it is more likely that the information-bearing photons emitted from the lamp at the back of the railroad car would excite the visual pigments of the observer before the information-bearing photons emitted from the lamp on the front of the railroad car excite the visual pigments of the observer.

**Conclusion**

In this paper I have described a commonsense and intuitive way to explain the velocity dependence of simultaneity in terms of changes in the spatial and temporal characteristics of electromagnetic waves that result from the new relativistic Doppler effect. That is, while the speed of light is isotropic and invariant for all observers; as a consequence of the Doppler effect-induced time lag, the propagation of the spatial ($k$ and $\lambda$) and temporal ($\omega$ and $\nu$) characteristics of light as well as its momentum ($\hbar k$), energy ($\hbar \omega$) and information content is anisotropic. By using a physically-meaningful theory of measurement that takes into consideration the new relativistic Doppler effect equation and the angle-dependent time lag it introduces, all inertial observers would be able to agree when two distant events occurred. The realistic interpretation of the relativity of simultaneity presented here contrasts with the unintuitive interpretation given by the Special Theory of Relativity.

The new relativistic Doppler effect equation presented here is a generalization of Einstein's [8] relativistic Doppler effect equation, which is limited to the special case of an "infinitely distant source of light" where $\cos^2 \theta$ is unity by definition. The quantitative differences predicted by the form-invariant to the second order new relativistic Doppler effect equation and Einstein's relativistic Doppler effect equation are testable using spectroscopic techniques. Such an experiment will simultaneously test whether the relativity of simultaneity is best explained by Einstein's Special Theory of Relativity, which explains the velocity dependence in terms of the relativity of space and time [30, 31], or by the velocity-dependent changes in the spatial and temporal characteristics of electromagnetic waves. While this paper is primarily concerned with the kinematic consequences of the Doppler effect, I have also given an account of the dynamic consequences of the Doppler effect that are also testable [32, 33].

**Acknowledgements**

I would like to dedicate this paper to my late colleagues Mark Jaffe and Carl Leopold, whose support, money could not buy.

**Appendix A. The New Relativistic Wave Equation and the Derivation of the Relativistic Doppler Effect Equation**

Assume that the following relativistic wave equation, which is form-invariant to the second order in all inertial frames, is the equation of motion that describes the properties of light observed by an observer in an inertial frame moving at velocity $v$ relative to the inertial frame of the light source:

$$\frac{\partial^2 \Psi}{\partial t^2} = cc' \frac{\sqrt{c + v \cos\theta}}{\sqrt{c - v \cos\theta}} \nabla^2 \Psi \tag{A1}$$

In the equation above, $\theta$ is the angle between the velocity vector and the angular wave vector pointing from the source to the observer. Assume that the following equation is a general plane wave solution to the second order relativistic wave equation given above:

$$\Psi = \Psi_o e^{i(\mathbf{k}_{observer} \cdot \mathbf{r} - \omega_{source} \frac{\sqrt{c + v \cos\theta}}{\sqrt{c - v \cos\theta}} t)} \tag{A2}$$

The general plane wave solution assumes that the direction of $\mathbf{r}$, which extends from the source to the observer, is arbitrary but $\mathbf{k}_{observer}$ is parallel to $\mathbf{r}$. Thus $\theta$ is the angle between the velocity vector and the angular wave vector. We can obtain the form-invariant to the second order relativistic dispersion relation by substituting equation A2 into equation A1 and taking the spatial and temporal partial derivatives:

$$cc' \frac{\sqrt{c + v \cos\theta}}{\sqrt{c - v \cos\theta}} i^2 k_{observer}^2 \Psi = i^2 \omega_{source}^2 \frac{c + v \cos\theta}{c - v \cos\theta} \Psi \tag{A3}$$

After canceling like terms, we get:

$$cc' k_{observer}^2 = \omega_{source}^2 \frac{\sqrt{c + v \cos\theta}}{\sqrt{c - v \cos\theta}} \tag{A4}$$

Since $c' = \dfrac{\omega_{source}}{k_{observer}}$, the above equation simplifies to:

$$c\, k_{observer} = \omega_{source} \frac{\sqrt{c + v \cos\theta}}{\sqrt{c - v \cos\theta}} \tag{A5}$$

Solving for c, the speed of the wave, we get the relativistic dispersion relation:

$$c = \frac{\omega_{source}}{k_{observer}} \frac{\sqrt{c + v \cos\theta}}{\sqrt{c - v \cos\theta}} = 2.99 \times 10^8 \text{ m/s} \tag{A6}$$

The relativistic dispersion relation tells us that while the observed angular wave number varies in a velocity-dependent manner, the speed of light is invariant and always travels from the source to the observer at velocity c. That is, the relative velocity between the source and the observer "stretches" or "compresses" the amplitude of the light wave without changing its speed.

Letting $k_{observer} = \omega_{observer}/c$, we get a relativistic Doppler effect equation in terms of angular frequency:

$$\omega_{observer} = \omega_{source} \frac{\sqrt{c + v\cos\theta}}{\sqrt{c - v\cos\theta}} = \omega_{source} \frac{1 + \frac{v}{c}\cos\theta}{\sqrt{1 - \frac{v^2 \cos^2\theta}{c^2}}} \tag{A7}$$

Other forms of this relativistic Doppler effect equation can be obtained using the following substitutions: $\omega = 2\pi\nu = kc = 2\pi c/\lambda$. The relativistic Doppler equation tells us that even though the speed of the wave is invariant, the Doppler effect results in the introduction of a velocity-dependent time lag so that the time in which the amplitude and thus the information content of the wave reaches an observer is velocity dependent.

**Appendix B. Independent Derivation of the Relativistic Doppler Effect Equation**

The Lorentz transform used by Einstein [8] is sufficient but not necessary to mathematically model the relativistic Doppler effect first observed by Ives and Stillwell [15]. By comparing the derivation of the relativistic Doppler effect equation given by Einstein, Mermin [34] and Moriconi [35] with the derivation given below, one sees that the form of the unknown function that describes the velocity-dependence of the spectral properties of the observed light is not unique but depends on the initial *ansatz* (eq. 2.1 given in Moriconi or eq. B2 given below). The *ansatz* equations of the Special Theory of Relativity assume that the first-order velocity-dependent spectral properties of the observed light depend on the component of the relative velocity of the source or observer parallel to the angular wave vector while the second-order velocity-dependent spectral properties of the observed light depend exclusively on the magnitude but not the direction of the relative velocity of the source or observer. This is because Einstein [8] derived the relativistic Doppler effect equation after making the assumption that the second-order effect applied only to an "infinitely distant source of light," where $\cos^2\theta$ is equal to unity. This velocity is not relative in the true sense of the word since, if $\cos^2\theta$ is not equal to unity, the velocity cannot be completely determined by an observer localized at a given coordinate but only by an omniscient observer. Einstein's [8] relativistic Doppler effect equation is typically used as a general equation without taking into consideration the assumption of an "infinitely distant source of light" he used to derive it. By contrast, my *ansatz* equation is more general in that it does not assume an "infinitely distant source of light" but rather that the first-order and second-order velocity-dependent spectral properties of the light depend on both the magnitude and direction of the velocity vector–specifically on the component of the velocity vector parallel to the wave vector.

In this appendix, I will justify the form of the unknown function (ϕ) mathematically by deducing its form and symmetry without using the new relativistic wave equation. The resulting form of the unknown function is justified physically since it accounts for the results of the Ives-Stillwell experiments.

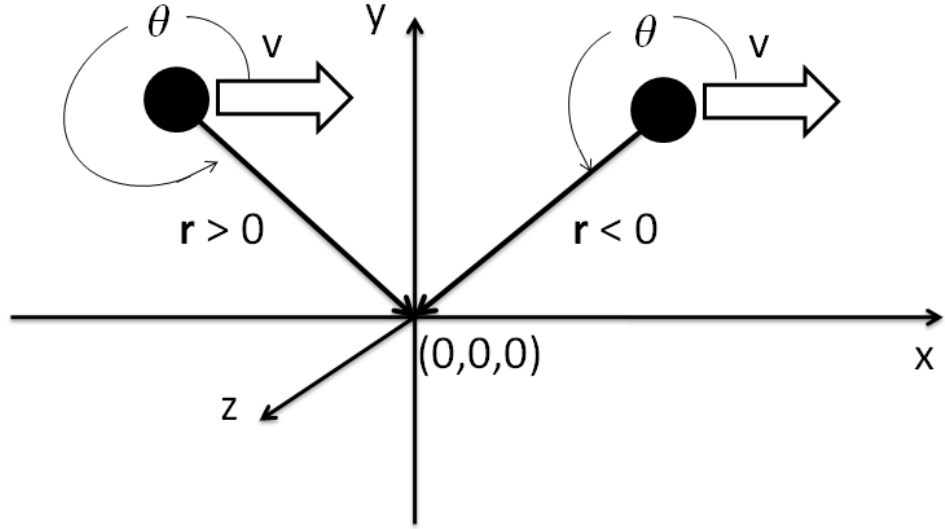

Fig. B1. A diagram showing two sources moving at velocity (**v**) > 0 relative to an observer at the origin (0,0,0). The figure could represent the two lamps on a railroad car moving relative to an observer on a platform. The vector **r** extends from the source to the observer. While the orientation of **r** is arbitrary, the angular wave vector **k** must travel parallel to **r** to use the new relativistic Doppler effect equations to determine the observed angular wave number. The orientation of **r** is given by the angle $\theta$ that originates parallel to **v**. An observer sees a blue-shifted source moving toward him or her when **v** and **r** point generally in the same direction. An observer sees a red-shifted source moving away from him or her when **v** and **r** point generally in the opposite direction.

Consider a light source moving relative to an observer at the origin of a Cartesian coordinate system (Figure B1). The angular wave number of the light observed ($k_{observer}$) will depend on the angular wave number of the source ($k_{source}$), the component of the relative velocity of the source (**v(r)**) parallel to the vector extending from the source to the observer (**r**), and the speed of light (c). Assuming that the angular wave vector **k** is parallel to **r**, the observed angular wave number is related to the angular wave number of the source by the following equation:

$$k_{observer} = k_{source}\, \phi\left(\frac{\pm v(r)}{c}\right) \tag{B1}$$

where, in Cartesian coordinates, **r** is the vector that points from the source to the observer and $\phi$ is an unknown function to be determined. While the orientation of **r** is arbitrary, assume that the angular wave vector **k** in question is parallel to **r** and that $\theta$ is the angle between **v** and **r**. When the dot product of **v** and **r** is positive, the source and observer are approaching each other and when the dot product of **v** and **r** is negative, the source and the observer are receding from each other.

Consider a source and observer moving relative to each other in an arbitrarily-oriented Cartesian coordinate system so that the velocity vector is parallel to the *x*-axis. Assuming the constancy of the speed of light (c), we get the following *ansatz* equation:

$$c = \frac{\sqrt{1 + \frac{v\cos\theta}{c}}}{\sqrt{1 - \frac{v\cos\theta}{c}}} \frac{ck_{source}}{k_{observer}} = \frac{\sqrt{1 + \frac{v\cos\theta}{c}}}{\sqrt{1 - \frac{v\cos\theta}{c}}} \frac{ck_{source}}{k_{source} \phi(\frac{\pm v(r)}{c})} \quad (B2)$$

When the source and the observer move toward each other ($\frac{\pi}{2} \geq \theta \geq \frac{3\pi}{2}$) for **v** > 0, eq. B2 becomes:

$$c = \frac{\sqrt{1 + \frac{v|\cos\theta|}{c}}}{\sqrt{1 - \frac{v|\cos\theta|}{c}}} \frac{ck_{source}}{k_{source} \phi(\frac{+v(r)}{c})} \quad (B3)$$

When the source and the observer move away from each other ($\frac{\pi}{2} \leq \theta \leq \frac{3\pi}{2}$) for **v** > 0, eq. B2 becomes:

$$c = \frac{\sqrt{1 - \frac{v|\cos\theta|}{c}}}{\sqrt{1 + \frac{v|\cos\theta|}{c}}} \frac{ck_{source}}{k_{source} \phi(\frac{-v(r)}{c})} \quad (B4)$$

Dividing eq. B3 by eq. B4, we get:

$$\frac{1 + \frac{v|\cos\theta|}{c}}{1 - \frac{v|\cos\theta|}{c}} \frac{\phi(\frac{-v(r)}{c})}{\phi(\frac{+v(r)}{c})} = 1 \quad (B5)$$

When the source and the observer are in the same inertial frame, eq. B1 becomes:

$$k_{observer} = k_{source} \quad (B6)$$

Consequently, when **v** = 0, $\phi(0) = 1$. When there is no relative motion, it is also true that:

$$k_{observer} = k_{source} \, \phi(\frac{+v(r)}{c}) \, \phi(\frac{-v(r)}{c}) \quad (B7)$$

Thus,

$$\phi(\frac{+v(r)}{c}) \phi(\frac{-v(r)}{c}) = 1 \tag{B8}$$

and the function $\phi(\frac{+v(r)}{c})$ is equal to the reciprocal of $\phi(\frac{-v(r)}{c})$. Substituting eq. B8 into eq. B5, we get:

$$\phi^2(\frac{v(r)}{c}) = \frac{1 + \frac{v |\cos\theta|}{c}}{1 - \frac{v |\cos\theta|}{c}} \tag{B9}$$

After taking the square roots of both sides, we get a solution for the function for a source and observer moving toward each other ($\frac{\pi}{2} \geq \theta \geq \frac{3\pi}{2}$), when **v** > 0:

$$\phi(\frac{+v(r)}{c}) = \frac{\sqrt{1 + \frac{v |\cos\theta|}{c}}}{\sqrt{1 - \frac{v |\cos\theta|}{c}}} = \frac{1 + \frac{v}{c}|\cos\theta|}{\sqrt{1 - \frac{v^2 \cos^2\theta}{c^2}}} \tag{B10}$$

Similarly, we get a solution for the function for a source and observer moving away from each other ($\frac{\pi}{2} \leq \theta \leq \frac{3\pi}{2}$), when **v** > 0:

$$\phi(\frac{-v(r)}{c}) = \frac{\sqrt{1 - \frac{v |\cos\theta|}{c}}}{\sqrt{1 + \frac{v |\cos\theta|}{c}}} = \frac{1 - \frac{v}{c}|\cos\theta|}{\sqrt{1 - \frac{v^2 \cos^2\theta}{c^2}}} \tag{B11}$$

In order to emphasize the component of the velocity vector parallel to the wave vector, equations B10 and B11 can be combined into one equation:

$$\phi(\frac{\pm v(r)}{c}) = \frac{\sqrt{1 \pm \frac{v |\cos\theta|}{c}}}{\sqrt{1 \pm \frac{v |\cos\theta|}{c}}} = \frac{\sqrt{1 + \frac{v \cos\theta}{c}}}{\sqrt{1 - \frac{v \cos\theta}{c}}} \tag{B12}$$

Substituting eq. B12 into eq. B1, we get the relativistic Doppler effect equation for angular wave number:

$$k_{observer} = k_{source} \frac{\sqrt{1 + \frac{v \cos\theta}{c}}}{\sqrt{1 - \frac{v \cos\theta}{c}}} \tag{B13}$$

This form of the relativistic Doppler effect equation is identical with the form derived from the new relativistic wave equation in Appendix A. It differs from the usual form [8, 34, 35] of the relativistic Doppler effect equation because its derivation from the *ansatz* carries through the full vectorial nature of **r** and **v** to the second order. In the general case, we get:

$$k_{observer} = k_{source} \frac{\sqrt{1 + \frac{v(r)}{c}}}{\sqrt{1 - \frac{v(r)}{c}}} = k_{source} \frac{1 + \frac{v(r)}{c}}{\sqrt{1 - \frac{v(r)^2}{c^2}}} \tag{B14}$$